\documentclass[superscriptaddress,reprint,showpacs,amsmath,amssymb,aps,prl,floatfix]{revtex4-1}

\usepackage{bbold}
\usepackage{graphicx}% Include figure files
\usepackage{dcolumn}% Align table columns on decimal point
\usepackage{bm}% bold math
\usepackage{epstopdf}
\usepackage{ulem}
\usepackage{color}

\newcommand{\ket}[1]{$|#1\rangle$}

\newcommand{\nd}{Nd$^{3+}$}
\newcommand{\er}{Er$^{3+}$}

\newcommand{\YSO}{Y$_2$SiO$_5$}

\newcommand{\YAP}{YAlO$_3$}

\begin{document}

%\preprint{PG}

%%\title{}% 
%\title{A Quantum Memory for Microwave Photons with Rare Earth Nuclear Spins}%
\title{Coherent Storage of Microwave Excitations in Rare Earth Nuclear Spins}

\author{Gary Wolfowicz}
\affiliation{London Centre for Nanotechnology, University College London, London WC1H 0AH, UK }%
\affiliation{Dept.\ of Materials, Oxford University, Oxford OX1 3PH, UK}  
\author{Hannes Maier-Flaig}
\email{Present address: Walther-Meissner-Institut, Bayerische Akademie der Wissenschaften, 85748 Garching, Germany}
\affiliation{London Centre for Nanotechnology, University College London, London WC1H 0AH, UK }%
\author{Robert Marino}
\affiliation{PSL Research University, Chimie ParisTech - CNRS, Institut de Recherche de Chimie Paris, 75005, Paris, France}
\affiliation{LASIR CNRS UMR 8516, Universit\'e de  Lille, France}
\author{Alban Ferrier}
\affiliation{%
PSL Research University, Chimie ParisTech - CNRS, Institut de Recherche de Chimie Paris, 75005, Paris, France}
\affiliation{
Sorbonne Universit\'es, UPMC Univ Paris 06, 75005, Paris, France
 }%
\author{Herv\'e Vezin}
\affiliation{LASIR CNRS UMR 8516, Universit\'e de  Lille, France}
\author{John J. L. Morton }
\affiliation{%
London Centre for Nanotechnology, University College London, London WC1H 0AH, UK
 }%
 \affiliation{%
Dept.\ of Electronic and Electrical Engineering, UCL, London WC1E 7JE, UK
 }%
\author{Philippe Goldner}
\email{philippe.goldner@chimie-paristech.fr}
\affiliation{%
PSL Research University, Chimie ParisTech - CNRS, Institut de Recherche de Chimie Paris, 75005, Paris, France}%

\date{\today}% It is always \today, today,

\begin{abstract}
Interfacing between various elements of a computer --- from memory to processors to long range communication --- will be as critical for quantum computers as it is for classical computers today. Paramagnetic rare earth doped crystals, such as \nd:\YSO (YSO), are excellent candidates for such a quantum interface:  they are known to exhibit long optical coherence lifetimes (for communication via optical photons), possess a nuclear spin (memory) and have in addition an electron spin that can offer hybrid coupling with superconducting qubits (processing). 
Here we study two of these three elements, demonstrating coherent storage and retrieval between  electron and $^{145}$Nd nuclear spin states in \nd:YSO. We find nuclear spin coherence times can reach 9 ms at $\sim$5~K, about two orders of magnitude longer than the electron spin coherence, while quantum state and process tomography of the storage/retrieval operation reveal an average state fidelity of 0.86. The times and fidelities are expected to further improve at lower temperatures and with more homogeneous radio-frequency excitation.
%These results further emphasise that rare earth doped crystals have the potential to provide a powerful quantum interface and motivate future studies combining strong coupling with optical and microwave cavities. <-- Probably a better place for this is in the conclusions, the abstract is already too long***

%Coherent transfer of quantum states between microwave photons and spin ensembles can provide long lived memories to superconducting qubits. Here, we report storage and retrieval of microwave photons in a paramagnetic rare earth doped crystal. By transferring the electron spin coherence to a nuclear transition, storage times can reach up to 9.2 ms, about two orders of magnitude longer than the electron spin coherence lifetime. Quantum state and process tomography  have been investigated, resulting in fidelities of 0.86 and 0.63 respectively, limited by  radio-frequency field inhomogeneities and electron spin dephasing. These results show that rare earth doped crystals, which are known to exhibit long optical coherence lifetimes,  could provide a long lived  quantum interface between  superconducting qubits photons in the microwave and optical range. 

\end{abstract}

% PACS, the Physics and Astronomy                             % Classification Scheme.
\maketitle

%{\color{blue}
%1. Hybrid quantum systems composed of spin ensembles strongly coupled to superconducting resonators and circuits have recently emerged as powerful paradigm, perhaps offering the hours-long coherence times of nuclears spins as a resource for superdoncuting qubits
%
%2. Such circuit QED experiments mirror the experiments in optical quantum memories performed with atomic ensembles and optical cavities
%
%3. Bringing both of these together is a very attractive idea, as it would enable a truly versatile quantum interface, connecting quantum memory, processing and communication and allowing faithful conversion of microwave to optical photons.
%}

Hybrid quantum systems composed of spin ensembles strongly coupled to superconducting resonators have recently emerged as a promising route for quantum memories operating in the microwave regime~\cite{Julsgaard:2013br,Afzelius:2013ga}. Such memories offer the possibility exploiting electron spin coherence times of up to seconds~\cite{Wolfowicz:2013ix} as a resource for superconducting qubits, whose coherence times so far extend only to tens of microseconds~\cite{Rigetti:2012en}. Strong coupling has been observed between superconducting resonators and various paramagnetic impurities, including NV centres in diamond~\cite{Kubo:2010iq} and erbium ions in \YSO\ (YSO) and \YAP~\cite{Probst:2013hn,Tkalcec:2014dy} leading to reversible coherent storage of (large numbers of) microwave photons within spin ensembles~\cite{Kubo:2012bn,Zhu:2011js}. These paramagnetic impurities are often coupled to nuclear spins which can offer a further resource for storage --- electronic spin coherences can be transferred to and from a nuclear spin \cite{Morton:2008et,Brown:2011jm,Wu:2010fm}, to access coherence times as long as hours~\cite{Saeedi:2013dy}.
%,Zhong14}. %% Note this paper is about to come out in Nature any week now, so I suggest we just re-insert the reference when it does.

The proposal to use solid-state spin ensembles as microwave quantum memories is in many ways inspired by results on using impurities in solid for optical quantum memories~\cite{Tittel:2009bp,Goldner:2015ve} where optical excitations are stored in rare earth (RE) nuclear spins~\cite{Longdell:2005ik,Afzelius:2010fh,PhysRevLett.111.020503,Heinze:2013dh}. Very long storage time can be expected, as nuclear spin coherence lifetimes in such materials extend up to 6 hours. Entanglement storage \cite{Clausen:2011uw} and light-matter teleportation at telecom wavelength \cite{Bussieres:2014dc} have been demonstrated in \nd:\YSO, for which optical coherence lifetimes of 90 $\mu$s have been measured~\cite{Macfarlane:2002ug,Usmani:2010hd}. 
%In \eu:YSO, nuclear spin coherence lifetimes in such materials extend up to 6 hours.%.~\cite{Zhong14}. %% Note this paper is about to come out in Nature any week now, so I suggest we just re-insert the reference when it does.

Bringing together both optical and microwave strong coupling techniques on the same ensemble would enable a versatile quantum interface, connecting quantum memory, processing and communication and potentially allowing faithful conversion of microwave to optical photons~\cite{OBrien:2014dc, Williamson:2014fb}. However,  hyperfine coherence lifetimes have so far only been studied for RE ions with an even number of $f$ electrons and no electron spin \cite{Louchet:2008cj,Alexander:2007fz,Fraval:2004cu,Goldner:2009bw}. It is therefore unknown whether nuclear spins could still provide a memory resource when the RE ions are paramagnetic, as required for coupling to microwave excitations.
In this Letter we study a paramagnetic RE doped crystal \nd:\YSO\ and measure electron and nuclear spin coherence times of up to 100~$\mu$s and 9.2 ms, respectively. We further demonstrate coherence transfer between electron and nuclear spin degrees of freedom in the \nd ion --- quantum state and process tomography show transfer fidelities above the classical limit. These results suggest that quantum memories for microwave photons with access to long storage times are achievable in rare earth doped crystals. %suggesting a way to build a light-matter quantum interface operating with both microwave and optical photons. 

% \cite{Wu:2010fm}
%However, this is about the longest reported values for RE ions electronic spins \cite{Bertaina:2007dm}. 

\begin{figure}
\begin{center}
\includegraphics[width=\columnwidth]{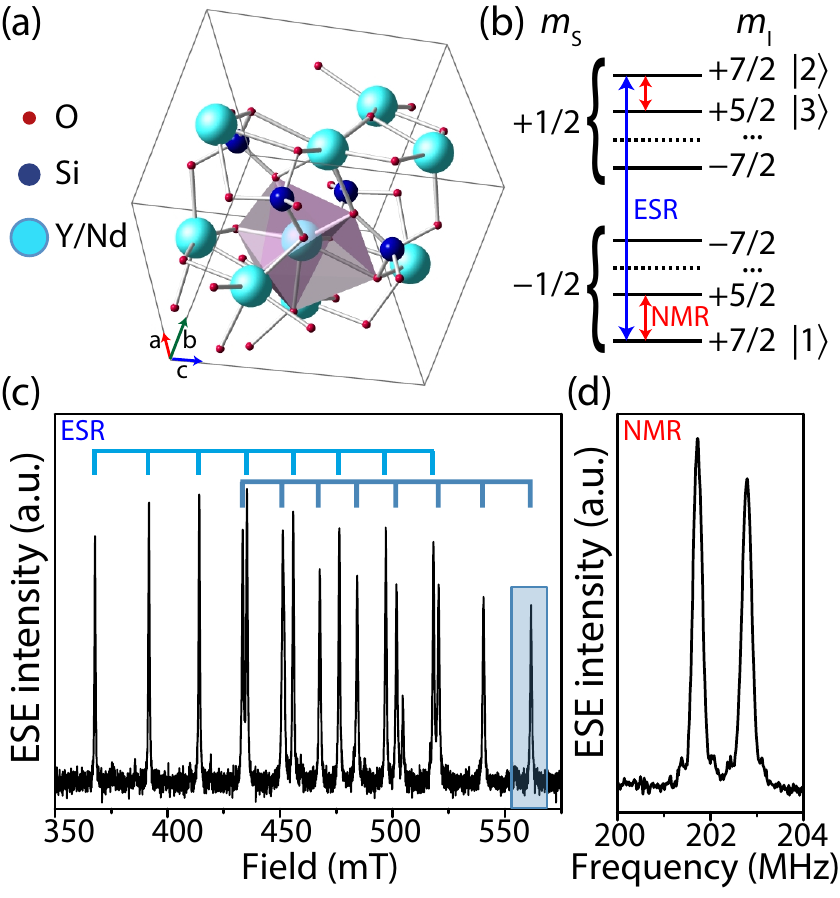}
\caption{(a) \YSO{} crystal structure showing the coordination polyhedron corresponding to \nd{} site within a unit cell. (b) Schematic energy level diagram of $^{145}$\nd{} ions highlighting relevant electron and nuclear spin transitions in blue and red, respectively. (c) 
Field swept ESE spectrum for a magnetic field close to D1 (T=6.5 K) showing two sets of ESR transitions from two magnetically inequivalent \nd{} classes. The electron spin transitions at 561.5 mT was used for all ENDOR experiments, such as the spectrum shown in (d).}
\label{spectra}
\end{center}
\end{figure}

\YSO\ (YSO) is a monoclinic crystal ($C^6_{2h}$ space group) with two crystallographic sites of $C_1$ symmetry for Y$^{3+}$ ions, which can be substituted by \nd{} ions (Fig. \ref{spectra}(a)).  Each site is divided in two classes related by a $C_2$ symmetry along the crystal $b$ axis. For magnetic fields parallel or perpendicular to the $b$ axis, ions in the two classes are magnetically equivalent. \nd{} has a [Xe]$4f^3$ electronic configuration, with a $^4$I$_{9/2}$  ground multiplet. In $C_1$ symmetry, the crystal field (CF) splits the $J$ multiplets into twofold degenerate levels. At low temperature, only the lowest doublet is populated and the system can be considered as  an effective $S=1/2$ spin.  
\nd{} has two isotopes with a $I=7/2$ nuclear spin, $^{143}$Nd and $^{145}$Nd, with respective natural abundance of 12.2 and 8.3 \%, as well as 5 isotopes with zero nuclear spin. To reduce the concentration of ions not involved in the storage experiments and potentially causing dephasing, an isotopically pure crystal boule of 0.001 at.\% $^{145}$Nd:YSO was grown by the Czochralski method. Samples of about 1.5 mm$^3$ were cut with faces perpendicular to the b, $D1$ and $D2$ principal axes of the optical indicatrix. Experiments were performed using an X-band (9.7 GHz) Bruker electron spin resonance (ESR) spectrometer (Elexys 580) equipped with a helium cryostat. Microwave (mw) $\pi$ pulses were 32 ns long and radio-frequency (rf) pulses about 3~$\mu$s.
%Radio-frequency (rf) pulses  were produced by an XXX generator and amplified by an XXXXX. %Continuous wave (CW) spectra were obtained at 8.5 K on a Bruker Elexys E500 spectrometer.  

Figure \ref{spectra}(c) shows the field swept electron spin echo (ESE) spectrum obtained for a  magnetic field oriented close to the $D1$ axis. The 16 intense lines correspond  to the allowed ESR transitions ($\{\Delta m_I,\Delta m_S\} = \{0,\pm1\}$) for the two magnetically inequivalent classes of one site. For some orientations of the magnetic field, weaker lines corresponding to $I=0$ isotopes in the same site were observed. These results suggest that \nd{} ions preferentially occupy one of the Y$^{3+}$ crystallographic sites.  The full linewidth at half maximum of the transition at 561.5 mT is 12 MHz, which is comparable to the narrowest linewidths measured in \er:YSO, recently used to demonstrate strong coupling to a superconducting resonator~\cite{Probst:2013hn}. All  ENDOR and relaxation experiments below were performed at  561.5 mT. The Zeeman $g$ and hyperfine $A$ tensors were determined from  CW spectra obtained by rotating the sample in planes containing the static magnetic field and perpendicular to the $D1,D2$ and $b$ axes. A least squares fit to the ESR line positions gives the principal values of the $g$ tensor: $g_x = 1.49$, $g_y = -0.98$, $g_z = -4.17$ with the Euler angles ($xzx$ convention) relating the principal axes to the crystal axes $D1,D2,b$: $\alpha=192^\circ$, $\beta = 39^\circ$ and $\gamma=183^\circ$. In the same reference axes, the principal values of $A$ and the corresponding Euler angles are: $A_x = 398$, $A_y = 0.1$, $A_z = 827$ MHz and $\alpha=154^\circ$, $\beta = 34^\circ$ and $\gamma=200^\circ$. As expected in low symmetry, the $g$ and $A$ tensors are highly anisotropic, but their principal axes are nearly parallel, as was observed for site 1 in \er:YSO \cite{GuillotNoel:2006fi}.

An electron-nuclear double resonance (ENDOR) spectrum was recorded (Fig. \ref{spectra}(d)), using a Davies ENDOR sequence with Tidy pulse~\cite{Davies:1974gu, Tyryshkin:2006iy}. 
%Here, a mw $\pi$ pulse first inverts the electronic spin population. A rf $\pi$ pulse is then applied and, on resonance, transfers electron spin population, which decreases the ESE signal. 
The two ENDOR lines located at 201.7 and 202.8 MHz have Gaussian shapes with linewidths of 235 and 248~kHz respectively. Simulations confirm that these correspond to +5/2:+7/2 transitions in $m_I$, where the lower (higher) frequency line corresponds to the $m_S=+1/2$ ($m_S=-1/2$) transition. The coherence storage experiments described below involve the three transitions labeled \ket{1}...\ket{3} as shown in Fig.~1(b).

\begin{figure}
\begin{center}
\includegraphics[width=\columnwidth]{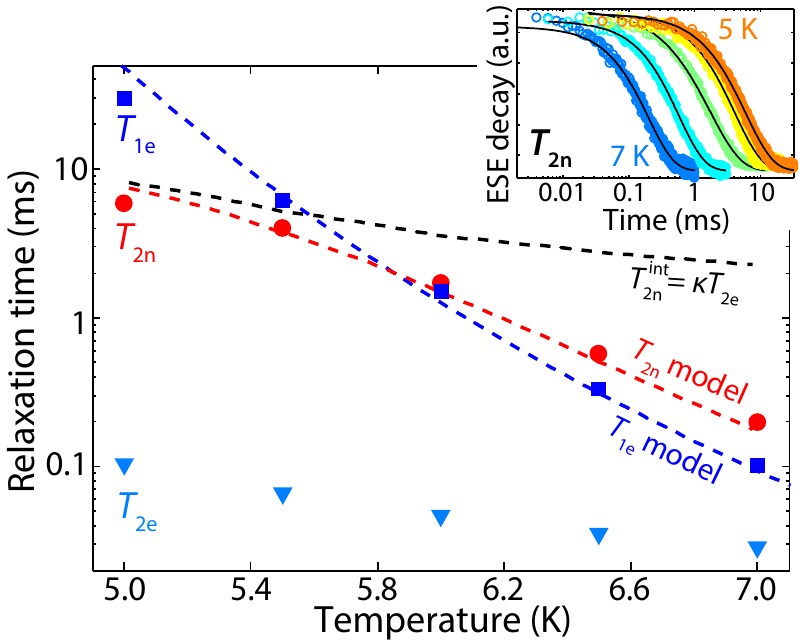}
\caption{Electron and nuclear spin relaxation times as a function of temperature: $T_{\rm 1e}$ (squares), $T_{\rm 2e}$ (triangles) and $T_{\rm 2n}$ (circles). In inset, corresponding decay curves for $T_{\rm 2n}$ from 5 to 7~K. $T_{\rm 1e}$ is modeled with an Orbach process $1/T_{\rm 1e} = A\exp({-\Delta E/k_{\rm B}T})$ with $A = 6 \times 10^{10}$ s$^{-1}$ and $\Delta E$ = 77 cm$^{-1}$ ($k_{\rm B}$ is the Boltzmann constant). $T_{\rm 2n}$ is limited by $2T_{\rm 2e}$ giving the relation $1/T_{\rm 2n} = 1/(2T_{\rm 1e}) + 1/T_{\rm 2n}^{\rm int}$, where $T_{\rm 2n}^{\rm int}$ is the decay time for the nucleus due to the spin environment only. As this decay has the same origin as for the electron, we can simply relate $T_{\rm 2n}^{\rm int} = \kappa T_{\rm 2e}$.}
\label{relaxation}
\end{center}
\end{figure}

The electron spin population relaxation time, $T_{1e}$, was measured by an inversion-recovery sequence as a function of temperature between 5 and 7 K.  $T_{1e}$ increases with decreasing temperature from 0.1 to 30 ms (Fig. \ref{relaxation}) and can be modeled above 5.5 K by an Orbach process with a CF level located 77 cm$^{-1}$ above the ground state, in reasonable agreement with the value of 88 cm$^{-1}$ deduced from optical measurements \cite{Beach:1990ee}. 
 The electron spin coherence lifetime $T_{2e}$ was also studied in the same temperature range (Fig. \ref{relaxation}), yielding stretched exponential decays with $T_{2e}$ in increasing from 28 to 106 $\mu$s with decreasing temperature. Stretched factors ranged between 1.2 and 1.5 below 6 K. We attribute the strong temperature dependence in  $T_{2e}$ to the effect of spectral diffusion resulting from interactions with a bath of electrons spins  undergoing spin-relaxation~\cite{Bottger:2006jo,Mims:1968vg}. When the bath relaxation rate  is much larger than the echo measurement scale, a stretch factor of $\approx 1.5$ indicates a Gaussian diffusion process \cite{Mims:1968vg}.  Using this model and taking \nd{} ions themselves as the spin bath, $T_{2e}$ can be estimated  from $T_{1e}$, the effective $g=1.5$ and \nd{} concentration ($9.4 \times 10^{16}$ ions/cm$^{3}$), which gives $T_{2e} = 471$ $\mu$s. This is is about four times longer than the measured value and can be explained by the anisotropy of the $g$ tensor, which increases the dipole-dipole interaction \cite{Maryasov:1982cg}. 
Angular variation in the $D1-D2$ plane showed that $T_{2e}$  is maximal in a region of about 5 degrees around $D1$ axis and decreases by a factor 2 at lower resonance fields.

\begin{figure}
\begin{center}
\includegraphics[width=\columnwidth]{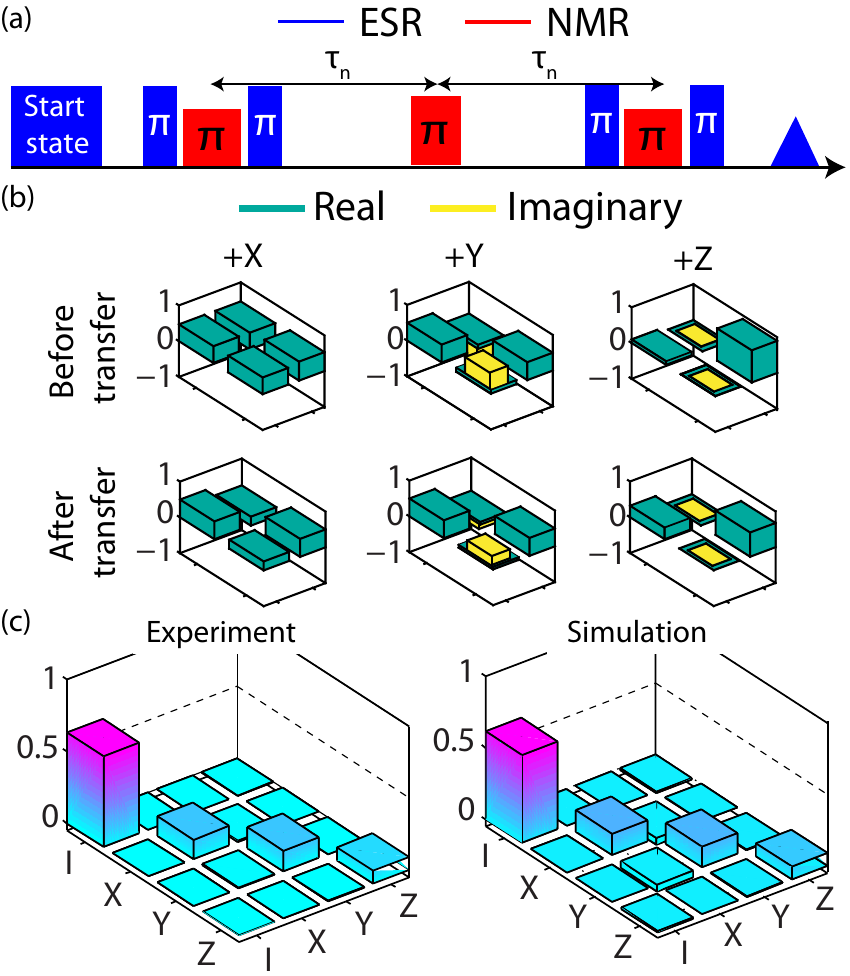}
\caption{(a) Sequence used for storing mw photons into nuclear spin coherences. (b) Input $+\sigma_X$, $+\sigma_Y$ and $+\sigma_Z$ (upper row) and corresponding output (lower row) density matrices are obtained by state tomography. Real and imaginary parts are shown in green and yellow, respectively. (c) Quantum process tomography matrix $\chi$ in the $(\mathbb{1},\sigma_X,\sigma_Y,\sigma_Z)$ basis. A perfect storage process would give only a $[\mathbb{1},\mathbb{1}]$ component. Left: Matrix reconstructed from experimental density matrices. Right: Simulated matrix considering pulse fidelity and spin relaxations.}
\label{storage}
\end{center}
\end{figure}

Transfer between electron and nuclear spin coherences was performed using the sequence shown in Fig.\ \ref{storage}(a) \cite{Morton:2008et}, which is fully compatible with the schemes designed for single photon operation \cite{Afzelius:2013ga,Julsgaard:2013br}. In our experiments, the memory input is a $\pi/2$ microwave pulse (consisting of ${\cal{O}}(10^{17}$) photons). It creates an electron spin coherence  on the \ket{1}:\ket{2} transition, which is then refocused by a $\pi$ pulse to remove the effect of inhomogeneous broadening. Before refocusing is complete, an rf $\pi$ pulse on the \ket{2}:\ket{3} transition transfers the coherence to \ket{1}:\ket{3}. At the time when this transition refocuses, a mw $\pi$ pulse transfers the electron spin coherence to the \ket{2}:\ket{3} NMR transition. To retrieve the coherent microwave signal from the nuclear spin ensemble, an rf $\pi$ pulse refocuses the \ket{2}:\ket{3} coherence, and then the sequence described above is applied in reverse order. A final mw $\pi$ produces an electron spin echo, which is the output of the memory. This scheme allows extending storage times beyond $T_{\rm 2e}$, limited instead by the nuclear spin coherence time $T_{\rm 2n}$.
%Efficiency of the storage protocol requires that pulse areas are precisely set. This was achieved by recording Rabi oscillations for the electron and nuclear spins. Fig. \ref{nucRabi} shows the oscillations  obtained on the (2)-(3) nuclear transition. In opposition to the results obtained on the electron spin (1)-(2) transition, a fast damping of the oscillations is observed, which we attribute to rf field inhomogeneities, since higher rf power did not change the damping rate. 

$T_{\rm 2n}$ was measured by monitoring the output echo amplitude as a function of $2\tau_n$ in the storage  sequence (Fig. \ref{storage}(a)). Echo decays were nearly exponential with maximal stretch factors of 1.25 and ranged from 184 $\mu$s at 7~K to 6~ms at 5~K (Fig. \ref{relaxation}). $T_{2n}$ is bounded by $2T_{1e}$ when there is significant hyperfine coupling~\cite{Morton:2008et}, and this limit is indeed observed for temperatures above 6~K. Below this temperature, some intrinsic nuclear spin decoherence mechanism is evident. We assume this intrinsic $T_{\rm 2n}$ follows the measured electron spin decoherence time $T_{\rm 2e}$, adjusted by some factor $\kappa$ to reflect the ratio of the effective $g$-factors for those ESR and NMR transitions: i.e.\ $1/T_{\rm 2n} = 1/(2T_{\rm 1e}) + 1/(\kappa T_{\rm 2e}$). $T_{\rm 2n}$ was found to depend significantly on the nuclear transition probed as well as on the static magnetic field orientation and ranged from 1.5 to 9.2 ms at 5~K, which can be understood by variations in $\kappa$.

The fidelity of the coherent storage and retrieval into the $^{145}$Nd nuclear spin subspace was characterised using quantum state tomography and quantum process tomography at 6.5 K to avoid low repetition rates due to the long $T_{1e}$. Quantum state tomography is performed by measuring the qubit state in the Pauli basis $(\sigma_X, \sigma_Y, \sigma_Z)$. Components $\sigma_X$ and $\sigma_Y$ are simply the real and imaginary part of the electron spin echo, while $\sigma_Z$ can be measured by an additional $\pi$/2 pulse immediately following the echo, to map $\sigma_Z$ onto $\sigma_X$~\cite{Morton:2008et}. To obtain the overall process matrix of the electron-nuclear-electron spin transfer, density matrices are measured for the set of electron spin input states: $\pm \sigma_X$, $\pm \sigma_Y$, $\pm \sigma_Z$ and $\mathbb{1}$ (Fig.\ \ref{storage}(b)). We are interested in obtaining the process matrix for the storage/retrieval operation itself, and so reference the output states against a simple two-pulse electron spin echo experiment with total duration equal to the time the coherent state resides in the electron spin degree of freedom, in the actual transfer sequence. 
%The input states for the process tomography are defined without the whole transfer sequence, and use a single mw $\pi$ at half time for obtaining the ESE. The output states are defined with the transfer sequence. 
In this way, losses related to electron spin relaxation, dephasing and state preparation are partly taken into account, but not errors related to the nuclear spin. The input and output states for the memory process are then linked by the relation, for a spin 1/2:
\begin{equation}
\epsilon(\rho_{\rm end}) = \sum_{\rm m,n=0}^3 \chi_{\rm mn} A_{\rm m} \rho_{\rm start} A_{\rm n}^\dagger
\end{equation}
where $\chi$ is the process matrix that is reconstructed, $A$ the operators from the Pauli basis $(\mathbb{1},\sigma_X,\sigma_Y,\sigma_Z)$ and $\rho_{\rm start}$ and $\rho_{\rm end}$ the input and output density matrices (for a particular electron spin initial state) \cite{Nielsen:2000vn,OBrien:2004cn}.

\begin{figure}
\begin{center}
\includegraphics[width=\columnwidth]{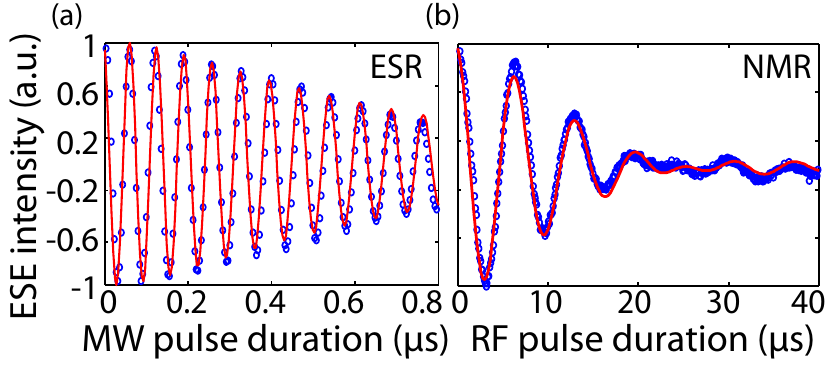}
\caption{(a) Electron Rabi oscillations obtained on the (1)-(2) transition (Fig. \ref{spectra}(b)). The decay is Gaussian, typical of mw pulse  inhomogeneity, with a standard deviation about 1.5\% of the Rabi frequency. (b) Nuclear Rabi oscillations obtained on the (2)-(3) transition.	The oscillations do not decay completely and can be modeled by the use of a truncated Gaussian for the rf pulse inhomogeneity. Such behaviour is likely due to the small sample size which only sees part of the magnetic field inhomogeneity (nearly 14\% here) from the rf coil.}
\label{ElecNucRabi}
\end{center}
\end{figure}

We measure an average state fidelity (where $F_{state} = \mathrm{Tr}(\sqrt{\sqrt{\rho_{\rm end}}\rho_{\rm start}\sqrt{\rho_{\rm end}}})^2$) of $\overline{F_{state}} = 0.86$, compared to what we would expect for an ideal memory, well above the classical limit of 2/3. The computed process matrix $\chi$ is shown in Fig. \ref{storage}(c) and we find a process fidelity $F_p=\mathrm{Tr}{(\chi \chi_{\rm ideal})}$ = 0.63, where $\chi_{\rm ideal}$ has just the identity component. Typically, average state and process fidelities are related by $\overline{F_{\rm state}} =( 2F_{\rm p} + 1)/3 = 0.75$ for a pure spin-1/2 \cite{Gilchrist:2005fw}. However, preparation and measurements are realized here on the electron spin, but conditional on a particular nuclear spin state, and so the reconstructed states do not span the full electron spin-1/2 state space. The reconstructed $\chi$ process matrix was well simulated using a Linblad master equation and taking into account electron and nuclear spin relaxation rates ($T_{\rm 1e}$, $T_{\rm 2e}$ and $T_{\rm 2n}$), as well as pulse inhomogeneities (Fig. \ref{storage}(c)). The latter were determined from fits to measured electron and nuclear spin Rabi oscillations (Fig. \ref{ElecNucRabi}). The main process errors can be assigned to two particular contributions: first, the low fidelity of the rf pulses results in the large components in the $[\sigma_X,\sigma_X]$ and $[\sigma_Y,\sigma_Y]$ part of $\chi$. Use of concatenated or adiabatic pulses would be expected to significantly address this issue \cite{Wimperis:1994uk,Genov:2014fm}. The second contribution is pure dephasing, as evidenced by the $[\sigma_Z,\sigma_Z]$ component in $\chi$, and is due to electron coherence decay during the application of the rf pulse. This could be significantly improved by lowering the temperature to increase $T_{\rm 2e}$ (Fig. \ref{relaxation}).  

In conclusion, we have shown that microwave excitations can be stored into a nuclear spin coherence in a isotopically pure rare earth doped crystal, $^{145}$\nd:\YSO. Storage times, determined by the nuclear coherence lifetime, can reach 9.2~ms, about two orders of magnitude longer than the electron spin $T_{\rm 2e}$ and the best superconducting qubit coherence times. Furthermore, these storage times could  be significantly increased by dynamical decoupling techniques \cite{Morton:2008et}. Given their long optical coherence lifetimes, our results show that paramagnetic rare earth doped crystals could be used as long lived quantum memories to interface superconducting qubits with both microwave and optical qubits. 

The authors thank M. Afzelius for useful discussions. 
The research leading to these results has received funding from the European Union's Seventh Framework Programme FP7/2007-2013/ under REA grant agreements no.\ 287252 (CIPRIS, People Programme-Marie Curie Actions), 247743 (QuRep) and 279781 (ASCENT), Idex  ANR-10-IDEX-0001-02 PSL$\star$ and R\'egion Nord-Pas-de-Calais. Work at UCL is supported by the EPSRC through a DTA and the Materials World Network (EP/I035536/1). J.J.L.M. is supported by the Royal Society.

%\bibliography{./Papers}


\end{document}